\documentclass{emulateapj}

\slugcomment{ApJ Letters, in press}

\shorttitle{AGN triggering and its luminosity dependence}
\shortauthors{Treister et al.}

\begin{document}

\title{Major Galaxy Mergers Only Trigger the Most Luminous AGN}

\author{E. Treister\altaffilmark{1}, K. Schawinski\altaffilmark{2,3,4,5}, C. M. Urry\altaffilmark{2,3,6}, B. D. Simmons\altaffilmark{2,3,6,7}}

\altaffiltext{1}{Universidad de Concepci\'{o}n, Departamento de Astronom\'{\i}a, Casilla 160-C, Concepci\'{o}n, Chile; etreiste@astro-udec.cl}
\altaffiltext{2}{Yale Center for Astronomy and Astrophysics, P.O. Box 208121, New Haven, CT 06520.}
\altaffiltext{3}{Department of Physics, Yale University, P.O. Box 208121, New Haven, CT 06520.}
\altaffiltext{4}{Einstein Fellow}
\altaffiltext{5}{Institute for Astronomy, Department of Physics, ETH Zurich, Wolfgang-Pauli-Strasse 16, CH-8093 Zurich, Switzerland}
\altaffiltext{6}{Department of Astronomy, Yale University, PO Box 208101, New Haven, CT 06520.}
\altaffiltext{7}{Department of Physics, University of Oxford, Keble Road, Oxford OX1 3RH, UK.}

\begin{abstract}
Using multiwavelength surveys of active galactic nuclei across a wide range of bolometric luminosities 
(10$^{43}$$<$$L_{bol}$(erg s$^{-1}$)$<$5$\times$10$^{46}$) and redshifts (0$<$$z$$<$3), we find a strong, redshift-independent 
correlation between the AGN luminosity and the fraction of host galaxies undergoing a major merger. That is, only the most luminous 
AGN phases are connected to major mergers, while less luminous AGN appear to be driven by secular processes. Combining this trend with AGN 
luminosity functions to assess the overall cosmic growth of black holes, we find that $\sim$50\% by mass is associated with 
major mergers, while only 10\% of AGN by number, the most luminous, are connected to these violent events. Our results suggest that to 
reach the highest AGN luminosities -where the most massive black holes accreted the bulk of their mass - a major merger appears to be 
required. The luminosity dependence of the fraction of AGN triggered by major mergers can successfully explain why the observed scatter 
in the M-$\sigma$ relation for elliptical galaxies is significantly lower than in spirals. The lack of a significant redshift 
dependence of the L$_{bol}$-$f_{merger}$ relation suggests that downsizing, i.e., the general decline in AGN and star formation activity 
with decreasing redshift, is driven by a decline in the frequency of major mergers combined with a decrease in the availability of gas at 
lower redshifts. 
\end{abstract}


\section{Introduction}

Theoretical models have shown that the energy output from a growing supermassive black hole can play a fundamental role in the star 
formation history  \citep{silk98,king03}. While it is clear now that most galaxies contain a supermassive black hole in their 
center, in only relatively few cases is this black hole actively growing.  This indicates that black hole growth is most likely episodic, with each luminous event 
lasting $\sim$10$^7$-10$^8$ years \citep{dimatteo05}. Hence, an obvious question is what triggers these black hole growth episodes?

Major galaxy mergers provide a good explanation, since as simulations show, they are very efficient in driving gas to the galaxy center 
\citep{barnes91}, where it can be used as fuel for both intense circumnuclear star formation and black hole growth. Indeed, a clear link 
between quasar (high-luminosity AGN) activity and galaxy mergers has been seen in intensely star-forming galaxies like Ultra-luminous
infrared galaxies (ULIRGs) and in some luminous quasars \citep[e.g.,][]{sanders88}. In contrast, many AGN are clearly not in mergers or 
especially rich environments \citep{de-robertis98}. Instead, minor interactions \citep{moore96}, instabilities driven by galaxy bars 
\citep{kormendy04} and other internal galaxy processes might be responsible for these lower activity levels. Understanding the role of 
mergers is further complicated by the difficulty of detecting merger signatures at high redshifts.

In order to reconcile these potentially contradictory observations it has been suggested that the AGN triggering mechanism
is a function of luminosity and/or redshift \citep[][and others]{finn01}. More recently, \citet{hopkins09a} used five indirect tests 
to conclude that the triggering mechanism is strongly luminosity-dependent and more weakly redshift-dependent, so that only the most 
luminous sources, which are preferentially found at $z$$>$2, are triggered by major mergers. Thanks to results from large
AGN surveys, which now include heavily-obscured IR-selected sources, and recent deep high-resolution observations carried out with 
the \emph{Hubble} WFC3 detector, it is now possible to obtain reliable morphological information even for high-$z$, low 
luminosity sources. In this paper we determine directly whether accreting black holes over a broad luminosity range are hosted by galaxies
undergoing a major merger. We use visual inspection to find tidal tails, prominent clumps of dust and/or star formation, or other indicators 
of recent major mergers that would suggest they could act as triggers for the black hole growth episode. Our sample covers over three
decades in luminosity and redshifts $z$$\sim$0-3 and thus removes the usual luminosity-redshift degeneracy in flux-limited samples.
Where needed, we assume a $\Lambda$CDM cosmology with $h_0$=0.7, $\Omega_m$=0.27 and $\Omega_\Lambda$=0.73.

\section{Analysis of Archival Observations}

To measure the fraction of AGN hosted by a galaxy undergoing a major merger as a function of luminosity and redshift, we compiled 
information from AGN samples selected from X-ray, infrared and spectroscopic surveys. X-ray surveys currently provide the most 
complete and cleanest AGN samples \citep{brandt05}; the deepest X-ray surveys performed with Chandra and XMM are sensitive up 
to column densities of $N_H$$\sim$10$^{23}$ \citep[e.g.,][]{treister04}, i.e., heavily obscured, nearly Compton-thick, 
sources. In order to properly sample the luminosity-redshift plane we follow the so-called ``wedding cake'' scheme, which combines
wide, shallow surveys that sample the rare and/or high-luminosity sources, with deep, narrow-field imaging that reaches high-redshift 
and/or low-luminosity AGN. Specifically, in this work we compile results from the $z$$\simeq$0 measurements of Swift/BAT-detected AGN 
\citep{koss10}, moderate luminosity AGN at $z$$\sim$1 in the COSMOS \citep{cisternas11}, AEGIS and GOODS \citep{georgakakis09} 
fields, and moderate luminosity AGN at $z$$\sim$2 in the CDF-S/CANDELS field \citep{schawinski11,kocevski12}. Despite the range of 
flux limits, this collective X-ray sample still shows the strong correlation between luminosity and redshift expected from flux-limited 
surveys. As a consequence, it becomes difficult to disentangle the possible effects of luminosity and redshift on the fraction of AGN 
triggered by galaxy mergers.

To widen the coverage of the luminosity-redshift plane and to increase the completeness of our sample by reaching more heavily 
obscured sources, we incorporate AGN candidates selected from infrared observations. In particular, we incorporate the IR-selected AGN 
at observed-frame 24~$\mu$m \citep{schawinski12} and 70~$\mu$m \citep{kartaltepe10b} in the CDF-S and COSMOS fields,
respectively. Finally, we complement these samples with AGN selected from the SDSS survey based on their narrow high-ionization emission lines 
\citep{kauffmann03a}, as reported by \citet{koss10}, and the high-luminosity optical and near-IR quasars studied by \citet{bahcall97} and \citet{urrutia08}, 
respectively. The samples used in this work are described in Table~\ref{table_surveys}.

\begin{deluxetable*}{llccccc}
\tablecolumns{7}
\tabletypesize{\scriptsize}
\tablecaption{Properties of AGN surveys studied in this work}
\tablehead{
\colhead{Survey} &\colhead{Morphological} & \colhead{Sources} & \colhead{Redshift} & \colhead{Luminosity} & \colhead{Merger} & \colhead{Symbol}\\
\colhead{} & \colhead{Classification} & \colhead{} &  \colhead{} & \colhead{10$^{44}$$L_{bol}$ erg s$^{-1}$} & \colhead{Fraction} & \colhead{}}
\startdata
\citet{veron-cetty91} & \citet{bahcall97}  & 20 & $z$$<$0.3 & 33-300 & 65\% & black triangle\\
\citet{glikman07} & \citet{urrutia08}  & 13 & 0.4$<$$z$$<$1 & 42-500 & 85\% & black square\\
\citet{kauffmann03a} & \citet{koss10}   & 72 & $z$$<$0.05 & 0.03-3 & 4\% & black circle\\
\citet{georgakakis09} & \citet{georgakakis09}  & 80 & 0.5$<$$z$$<$1.3 &  0.1-20 & 20\% & blue pentagon\\
\citet{tueller10} & \citet{koss10} & 72 & $z$$<$0.05 & 1-10 & 25\% & blue circle \\
\citet{hasinger07} & \citet{cisternas11}  & 140 & 0.3$<$$z$$<$1 & 2-20 & 15\% & blue square\\
\citet{luo08} & \citet{schawinski11}  & 23 & 1.5$<$$z$$<$3 & 0.1-20 & 9\% & blue cross\\
\citet{xue11} & \citet{kocevski12}  & 72 & 1.5$<$$z$$<$2.5 & 0.1-20 & 16.7\% & blue triangle\\
\citet{frayer09} & \citet{kartaltepe10b}  & 354 & 0.1$<$$z$$<$2 & 1-300 & 5-60\% & red triangles\\
\citet{treister09c} & \citet{schawinski12} & 28 & $z$$\sim$2 & $\sim$10 & 25\% & red square\\
\enddata
\label{table_surveys}
\end{deluxetable*}

The goal is to determine the physical mechanism(s) that provoked the AGN activity identified in these surveys. We separate sources into those 
that present evidence of external interactions (e.g., galaxy mergers) and those in which no signs of interactions are visible, based on the morphology 
of the AGN host galaxy. We expect that AGN triggered by external interactions will present distorted morphologies, obvious signs of interactions
and/or close neighbors. Morphologies of AGN host galaxies have been estimated before using automatic classifications  \citep[e.g.,][]{grogin05}, and 2-dimensional surface brightness 
fitting \citep{sanchez04}. However, as shown by simulations, using model-independent parameters such as CAS and the Gini coefficient in many cases can fail to identify
major mergers \citep{lotz08}, while surface brightness fitting typically involves assumptions of asymmetry that often break down in the case of major mergers, and additionally requires 
careful separation of nuclear light from host galaxy when used with AGN \citep{simmons08,koss11}. Hence, visual classification is still the most reliable option to determine if a galaxy is 
experiencing a major merger \citep{darg10}. Therefore, we focus here on surveys with visual merger classifications. For each survey we used the original classifications given
by the authors, which in most cases were determined by several independent visual inspections. We do not find evidence for significant differences in the classification
criteria from survey to survey.

\begin{figure*}
\begin{center}
\plottwo{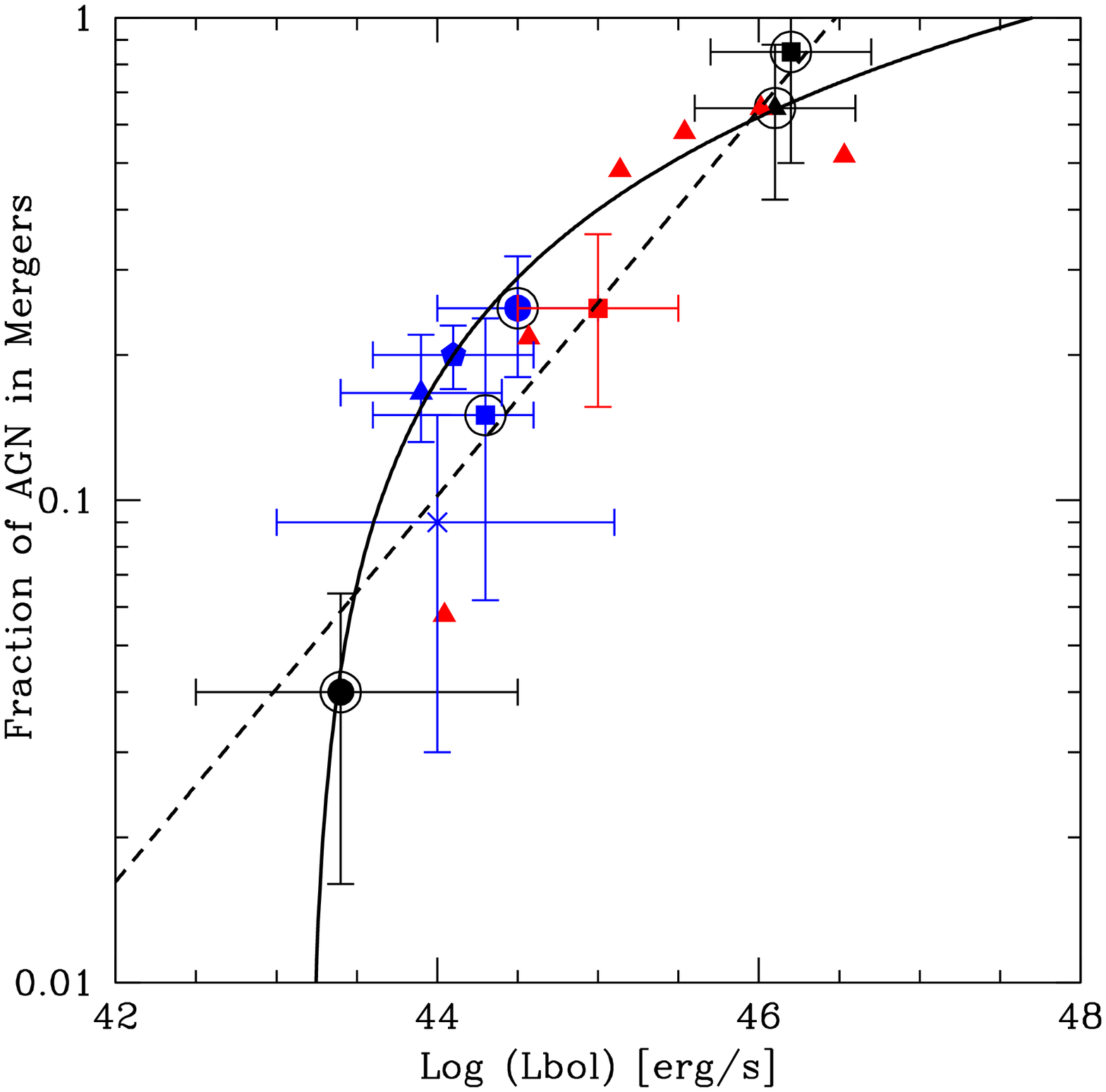}{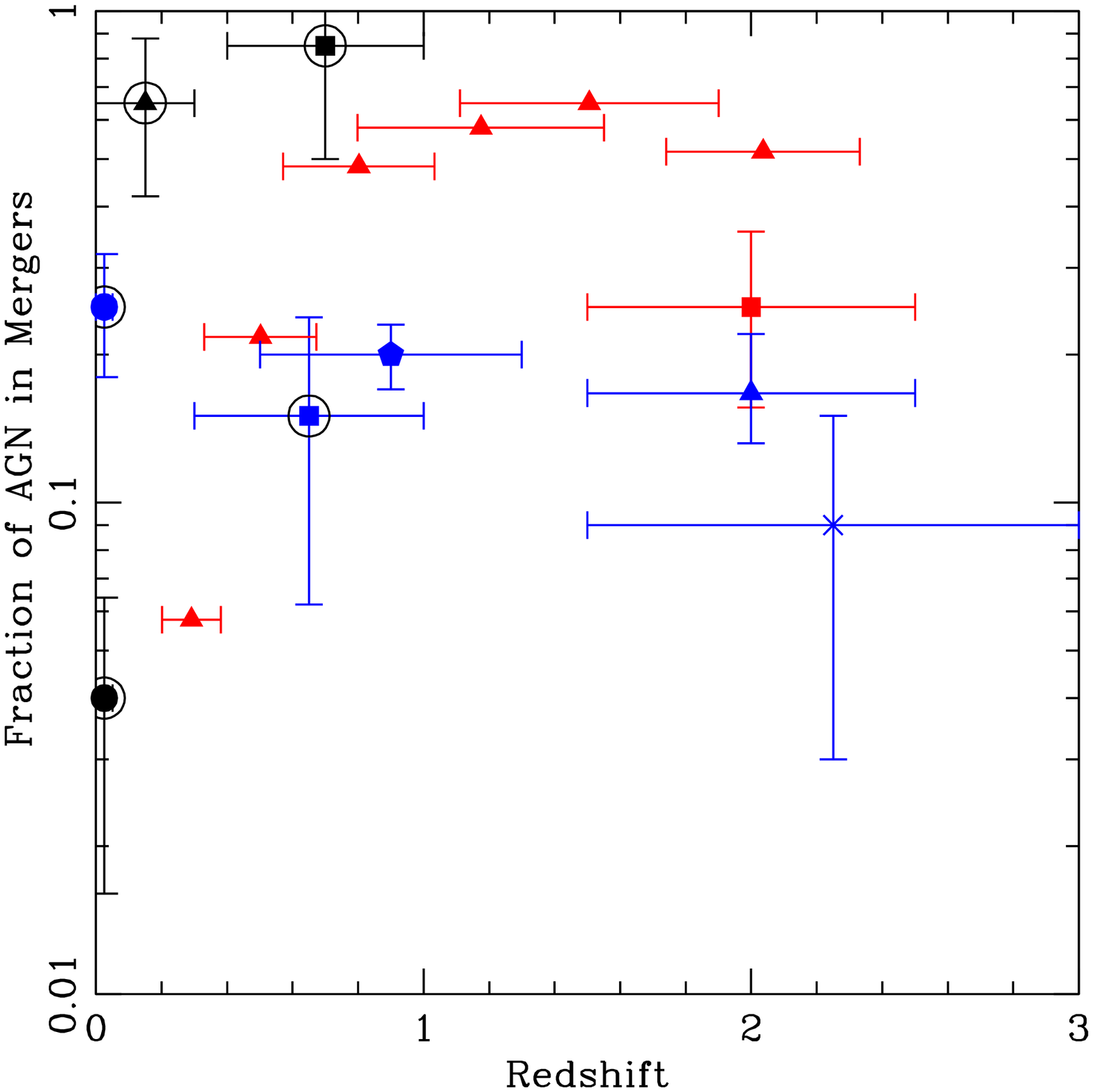}
\end{center}
\caption{{\it Left panel}: Fraction of AGN showing mergers as a function of the AGN bolometric luminosity. Colors indicate AGN selection method ({\it red}: infrared, {\it blue}: X-rays, {\it
black}: optical). Symbols used for each survey are presented in Table~\ref{table_surveys}. Encircled symbols show samples at $z$$<$1. {\it Solid line} shows a 
fit to the data assuming a linear dependence of the fraction on $\log$$(L_{bol})$, while the {\it dashed line} assumes a power-law dependence. {\it Right panel}: Fraction
of AGN showing mergers as a function of redshift.  There is a clear luminosity dependence, but no redshift dependence, suggesting that redshift is a second order effect in determining 
the dominant AGN triggering mechanism.}
\label{merg_frac_lum}
\end{figure*}

The fraction of AGN linked to galaxy mergers in these samples has been computed by dividing the number of AGN in which the host galaxy has been
classified as an ongoing merger or as having major disturbances by the total number of AGN. Figure~\ref{merg_frac_lum} shows the fraction of AGN showing mergers
as a function of bolometric luminosity, which increases rapidly, from $\sim$4\% at 10$^{43}$~erg~s$^{-1}$ to $\sim$90\% at 
10$^{46}$~erg~s$^{-1}$. We parametrize this dependence as a linear relation of the fraction of AGN showing mergers with luminosity,

\begin{equation}
\textnormal{frac}(L)=\frac{\log(L_{bol})-43.2}{4.5},
\end{equation}

or as a power law,

\begin{equation}
\textnormal{frac}(L)=\left(\frac{L_{bol}}{3\times10^{46} \textnormal{erg/s}}\right)^{0.4}.
\end{equation}

Both of these parameterizations, which provide good fits to the observed data, are shown in Figure~\ref{merg_frac_lum}. In order to assess the significance of this dependence, we compute the Pearson
correlation coefficient for frac$($$L$$)$ and $\log$$L$, finding a high value of $\sim$0.9, which indicates that a statistically strong linear correlation exist
between these two quantities. This correlation is also observed in individual samples, such as the infrared-selected sources in the COSMOS field \citep{kartaltepe10b}, confirming that
this correlation is not due to differences in classification schemes among different samples. We also test for possible correlations with other parameters. Interestingly, we do not find a significant correlation between 
redshift and the fraction of AGN in major mergers (Pearson correlation coefficient of $\sim$0.2), as is apparent in the right-hand panel of Figure~\ref{merg_frac_lum}. Nor do we find a significant correlation
between luminosity and redshift (Pearson coefficient $\sim$0.3), indicating that we are properly sampling the luminosity-redshift plane and thus high luminosity sources are not preferentially
at high redshifts. Our results suggest that while high-luminosity AGN triggered by major mergers are more common at high redshifts because 
of the increase of merger rate with redshift \citep[e.g.,][]{kartaltepe07}, this is not the dominant effect in determining the AGN triggering mechanism.

It seems unlikely that major mergers do not trigger AGN activity, and that it is instead caused by something else. Specifically, it was observed that the merger rate increases not only with redshift as reported 
above but also with galaxy mass \citep[e.g.,][]{hopkins10}. In contrast, the distribution of AGN luminosities is independent of stellar mass \citep{aird12}. Hence, we conclude 
that the observed increase in the fraction of AGN showing mergers with luminosity is directly linked to the triggering mechanism rather than to the galaxy mass. Furthermore, despite the fact that galaxy mergers 
are known to increase with redshift, our results indicate that AGN bolometric luminosity (and thus accretion rate) is the largest factor. Establishing the possible presence of other evolutionary effects in the AGN triggering mechanism would require, in addition, measuring the fraction of AGN in mergers relative to the total merger population, which is beyond the scope of the present work.

\section{Contribution of Black Hole Growth to the Extragalactic X-ray and UV Light}

We combine the observational results reported here with existing AGN luminosity functions in order to establish the importance of galaxy 
mergers in the growth of supermassive black holes in terms of the total accreted mass. In order to do this, we incorporate the linear 
parameterization of the luminosity dependence of the dominant AGN triggering mechanism (Equation 1) into the X-ray luminosity function and evolution 
of \citet{aird10}, combined with the distribution of obscuration and evolution of heavily obscured sources of \citet{treister09b,treister10}.
While the original work of \citet{treister09b} was based on the X-ray luminosity function of \citet{ueda03}, only small differences are found if 
the luminosity function of \citet{aird10} is used instead.  

\begin{figure}
\begin{center}
\includegraphics[angle=270,width=0.49\textwidth]{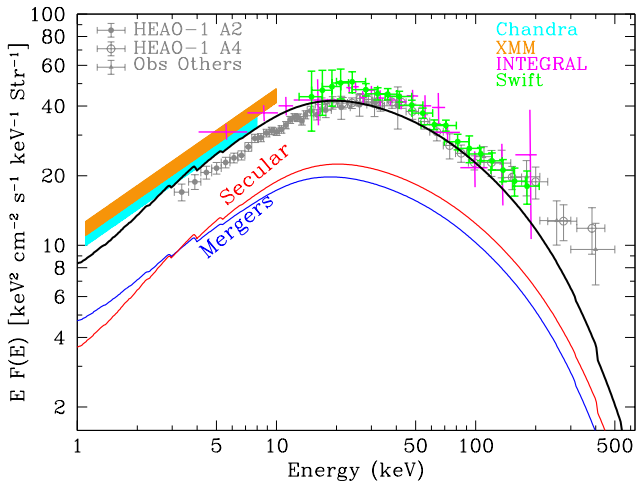}
\end{center}
\caption{Spectral energy distribution of the extragalactic X-ray ``background'' (which is actually the sum of AGN emission), as a function of observed-frame energy. Observational data points are 
summarized by \citet{treister09b}. Merger-triggered AGN ({\it blue line}) contribute roughly equal amounts of light as black hole growth ({\it red line}). Most of the X-ray background emission comes 
from $z$$<$1 \citep{treister09b}, hence the relative importance of secularly-triggered AGN. The extragalactic background light from higher redshift AGN peaks in the optical/UV and is dominated by 
luminous, merger-triggered AGN. The spectral shapes of the merger and secular contributions are slightly different since the fraction of obscured sources is a function
of luminosity.}
\label{xrb_mergers}
\end{figure}

The spectral shape and intensity of the extragalactic X-ray light (also known as the X-ray background) can tell us about the average properties of the AGN population. Using the models of \citet{treister09b}
with the AGN luminosity function of \citet{aird10} and the luminosity dependence of the fraction of AGN triggered by major mergers, we can estimate their contribution
to the background radiation in X-rays. In Figure~\ref{xrb_mergers} we show separately the contributions to the X-ray background from AGN triggered by secular processes 
and major mergers, which contribute nearly equally to the X-ray background. This is because most of the X-ray background comes from $z$$<$1 sources \citep[e.g.,][]{treister09b}, where AGN 
activity due to secular processes is relatively more important. This is particularly true at $E$$>$10~keV, where AGN emission is roughly unaffected by obscuration. Because of the luminosity dependence of the
fraction of obscured AGN \citep[e.g.,][]{ueda03}, AGN triggered by secular processes are relatively more obscured than those attributed to
major galaxy mergers, which explains the different spectral shapes in Figure~\ref{xrb_mergers} and the fact that AGN triggered by mergers are more important
at E$<$5 keV. We note that a population of high-luminosity heavily-obscured quasars likely associated with major mergers have been reported by \citet{treister10} and
others. These sources are mostly found at $z$$\sim$2 and show evidence of very high, Compton thick, levels of obscuration. Hence, these sources do not contribute 
significantly to the X-ray background radiation at any energy.

At UV wavelengths, the picture is quite complementary. The moderate luminosity AGN ($L_X$$\sim$10$^{43-44}$erg~s$^{-1}$ or $L_{bol}$$\sim$10$^{44-45}$erg~s$^{-1}$) tend to be obscured but 
Compton thin, so while bright in X-rays their rest-frame optical and UV light is completely absorbed. Meanwhile, high luminosity AGN tend to have strong UV emission, so they make up most of the 
extragalactic UV light; most are found at $z$$\sim$2 and are likely linked to the later stages of a major galaxy merger.

\section{Implications for Black Hole Growth}

\begin{figure*}
\begin{center}
\plottwo{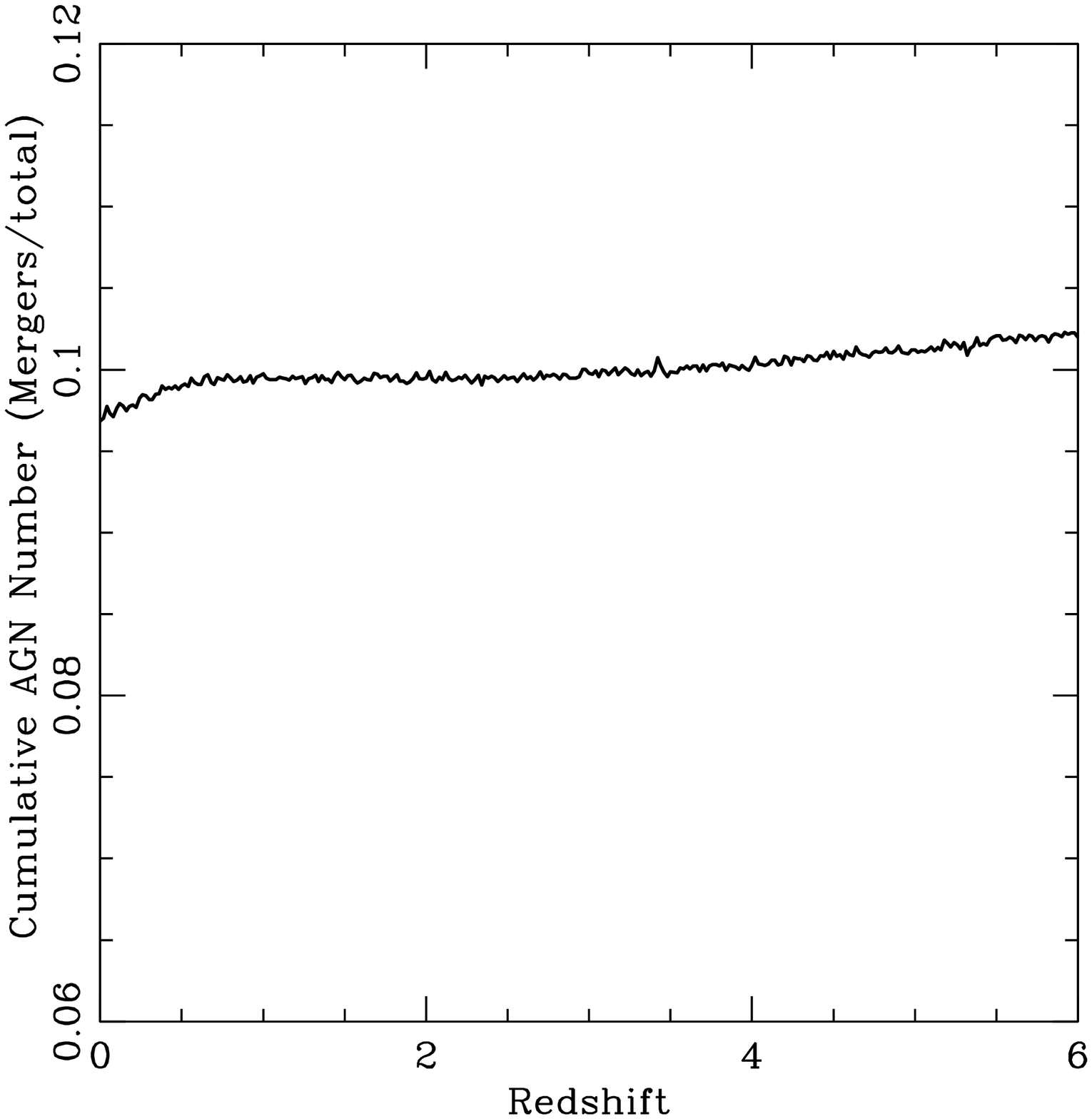}{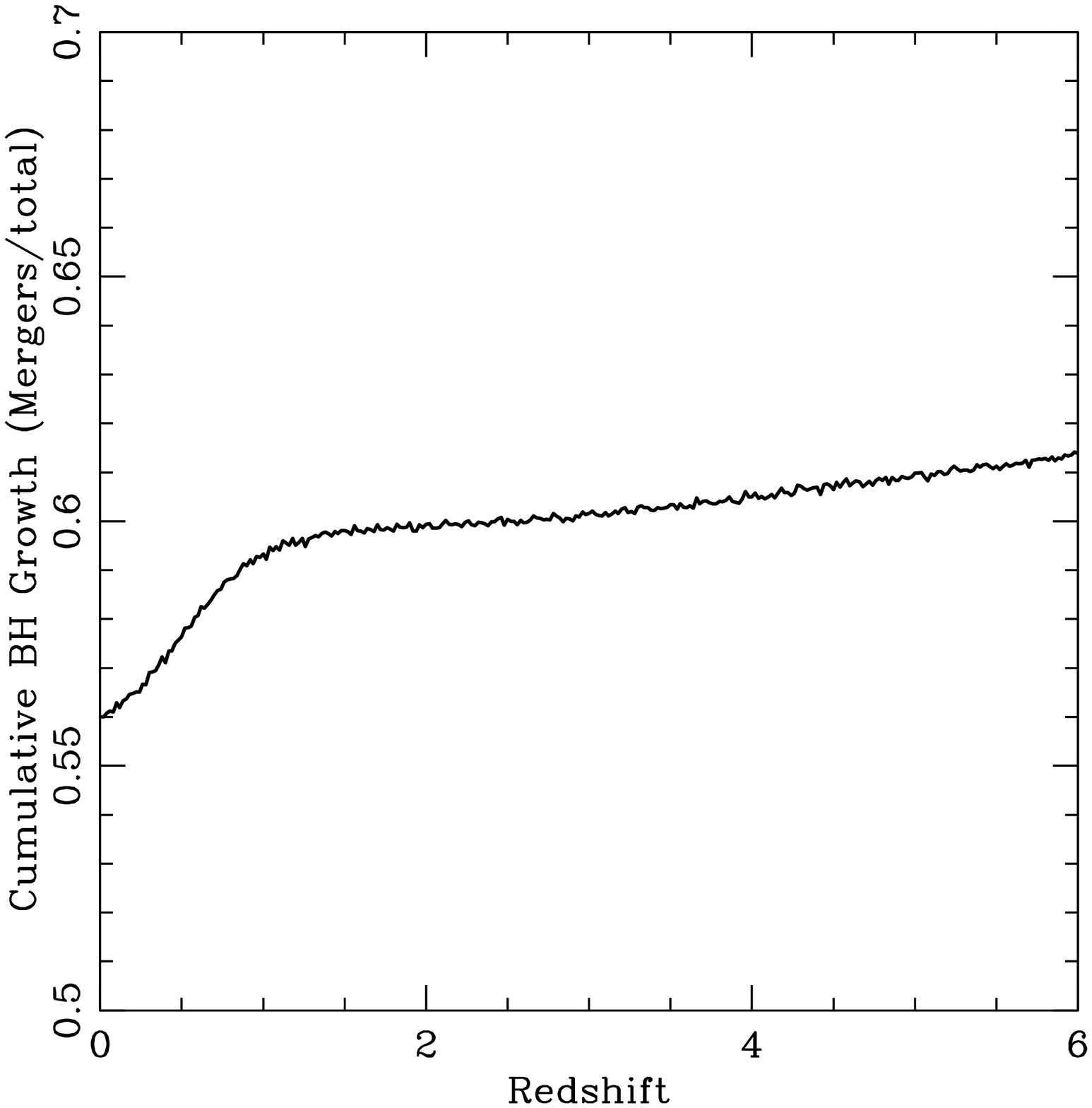}
\end{center}
\caption{{\it Left panel}: Cumulative number of merger-triggered AGN relative to the total number of AGN as a function of redshift.
While secular-triggered AGN vastly outnumber those triggered by major mergers, by about a factor of $\sim$10, the latter are on average significantly 
more luminous, thus explaining why they constitute $\sim$60\% of black hole accretion. {\it Right panel:} Cumulative fraction of black hole accreted mass in 
AGN triggered by mergers as a function of redshift, assuming a constant efficiency for converting mass to light. Black hole accretion is dominated by merger-triggered AGN at all redshifts but especially at $z$$>$1. At $z$$\sim$1, the much lower gas and merger fractions lead to a dominance of secular processes in AGN accretion.}
\label{red_dep}
\end{figure*}

In Figure~\ref{red_dep} we show, as a function of redshift, the amount of black hole growth and number of AGN triggered by major galaxy mergers relative to those associated
with secular processes. As can be seen and was previously reported \citep[e.g.,][]{treister10}, black hole growth occurs mostly in accretion 
episodes triggered by major galaxy mergers, although secular processes are still important. This is particularly true at $z$$\gtrsim$2, where there is $\sim$60\% more black hole growth
in merger-triggered AGN than in those growing via secular processes. At lower redshifts, there are relatively fewer galaxy mergers and so secular processes become slightly
more important. Furthermore, at lower redshifts dry mergers become more common than gas-rich major mergers \citep{kauffmann00}. Since the availability of gas is a critical
factor in determining the black hole accretion rate, this further explains why major mergers are relatively more important at high redshifts. It is interesting to note that the diminishing 
role of mergers coincides with the decline in the space density of black hole growth and with the observed decline in the cosmic star formation rate \citep{dahlen07}, i.e., cosmic downsizing. Integrated 
over the whole cosmic history, to $z$=0, 56\% of the total black hole growth can be attributed to major galaxy mergers.

In terms of numbers, the population is strongly dominated by secularly-triggered AGN. Indeed, as can be seen in Figure~\ref{red_dep}, $\sim$90\% of AGN
at all redshifts are associated with secular processes. This explains the conclusions of previous studies, mostly based on X-ray surveys \citep[e.g.][]{cisternas11,schawinski11,
kocevski12} of moderate luminosity AGN, which found that normal disk-dominated galaxies constitute the majority of the AGN host galaxies. So, we conclude that while most AGN
are triggered by secular processes, most of the black hole growth, particularly at high redshifts, can be attributed to intense accretion episodes linked to major
galaxy mergers.  

This calculation adopts the simplifying assumption that the accretion efficiency, average Eddington ratios and AGN duty cycle are constant and the same for secular and merger-triggered AGN. 
While this is obviously an idealized approach, it is justified by the good agreement between model results and observations of the integrated black hole mass function at $z$=0 \citep{treister09b}. 
Recently, \citet{draper12} reported a more sophisticated simulation, in which they assumed a distribution of Eddington ratios (and in general AGN light curves) for secular and merger-triggered 
AGN. Qualitatively, however, the results are very similar and consistent. 

While it is clear that at the major galaxy mergers are responsible for the highest luminosity events, it is expected that these AGN while eventually fade to lower luminosities. According to
model AGN light curves \citep[e.g.,][]{hopkins09b}, the nuclear luminosity can decrease by $\sim$4 orders of magnitude in $\sim$10$^8$ years. This is much shorter than the typical duration of
the merger sequence, $\sim$10$^9$ years \citep{dimatteo05}. Hence, it is unlikely that sources classified as secularly-triggered correspond to merger-triggered AGN in which the merger 
signatures are lost. Furthermore, and as shown by e.g., \citet{simmons11}, most low-luminosity AGN show significant disks, and thus they probably did not experience a relatively recent
major merger.

Our results so far refer only to the integrated black hole growth. Assuming that the Eddington ratio does not depend strongly on black hole mass \citep{woo02}, it is clear that the most massive
black holes are gaining most of their mass in episodes triggered by major mergers at relatively high redshifts, $z$$\gtrsim$1.5. Smaller black holes are either growing slowly at all redshifts in episodes 
not related to major mergers or are experiencing their first significant growth episodes at relatively low redshifts $z$$<$1. Once massive black holes have acquired most of their mass, they can continue
growing in low-luminosity systems at low Eddington rates which are most likely triggered by secular processes \citep{simmons11}. This is consistent with the general downsizing paradigm, in the sense
that the biggest black holes grow faster and are the first ones to accrete most of their mass, while lower mass black holes are formed later \citep[e.g.,][]{ueda03}. We also note there will be a decline, with 
decreasing redshift, in the effectiveness of major mergers at triggering black hole growth, as the gas supply is used up and dry mergers begin to predominate.

These results can also help explain the observed scatter in the M-$\sigma$ and M-L relations. Theoretically, it is expected that black hole growth triggered by major galaxy mergers leads
to a tighter M-$\sigma$ relation compared to growth due to stochastic processes \citep{hopkins09a}. This is because the former affects both black hole growth and bulge formation simultaneously.
Indeed, the scatter of the M-$\sigma$ relation for elliptical galaxies is observed to be significantly lower than for spiral galaxies \citep{gultekin09}. It is possible that the scatter depends more on black hole 
mass than host galaxy type, since ellipticals host more massive black holes than spirals (e.g., \citealp{gultekin09}), but this is also consistent with the expectations from major mergers, which should produce 
the largest growth spurts. 

In summary, we report a strong observational correlation between the AGN bolometric luminosity and the fraction of AGN hosted by galaxies undergoing 
major mergers. In contrast, we find no significant evidence for a correlation between this fraction and redshift. This strongly suggests that at all redshifts, vigorous accretion episodes are directly linked to 
major galaxy mergers, while less significant nuclear activity is most likely triggered by secular processes. Hence, just having
galaxies with large amounts of gas and dust is not enough to trigger intense black hole growth, and to reach the highest black hole
masses requires at least one quasar episode ignited by a galaxy merger.  These happen preferentially at high redshift, providing a natural explanation for the 
downsizing of black hole growth and star formation. We conclude that the triggering mechanism is the 
most relevant factor in determining the AGN luminosity and hence the black hole accretion rate. By incorporating this luminosity dependence 
into AGN population synthesis models we find that merger triggered AGN and those triggered by secular processes contribute 
roughly equally to the extragalactic X-ray background emission. While $\sim$90\% of the AGN by number are triggered by secular processes, 
$\sim$50-60\% of the total black hole growth is due to nuclear activity ignited by major galaxy mergers. 

\acknowledgements

We thank the anonymous referee and Michael Koss for providing very constructive comments. ET received partial support from Center of Excellence in Astrophysics and Associated Technologies (PFB 06) and from FONDECYT 
grant 1120061. Support for the work of KS was provided by the National Aeronautics and Space Administration through Einstein
Post-doctoral Fellowship Award Number PF9-00069 issued by the Chandra X-ray Observatory Center, which is
operated by the Smithsonian Astrophysical Observatory for and on behalf of the National Aeronautics Space Administration 
under contract NAS8-03060. BDS and CMU acknowledge support from NASA through grant HST-AR-12638.01-A from the Space Telescope Science 
Institute, which is operated by the Association of Universities for Research in Astronomy under NASA contract NAS 5-26555.
CMU acknowledges support from Yale University.

\end{document}